\documentclass[twocolumn,superscriptaddress,showpacs]{revtex4}
\usepackage{amsmath,amsthm,amssymb}
\usepackage{latexsym}
\usepackage{array}
\usepackage{amscd,graphicx}
\usepackage{bm,textcomp}
\usepackage[titletoc]{appendix}
\usepackage{epsfig}
\usepackage{epstopdf}

\begin{document}

\title{Coupling spin ensembles via superconducting flux qubits}
\author{Yueyin Qiu}
\affiliation{Department of Physics, Fudan University, Shanghai 200433,China}
\affiliation{Beijing Computational Science Research Center, Beijing 100084, China}
\author{Wei Xiong}
\affiliation{Department of Physics, Fudan University, Shanghai 200433,China}
\author{Lin Tian}
\affiliation{School of Natural Sciences, University of California, Merced, California
95343, USA}
\author{J. Q. You}
\email{jqyou@csrc.ac.cn}
\affiliation{Beijing Computational Science Research Center, Beijing 100084, China}
\date{\today }

\begin{abstract}
We study a hybrid quantum system consisting of spin ensembles and
superconducting flux qubits, where each spin ensemble is realized using the nitrogen-vacancy centers in a diamond crystal and the nearest-neighbor spin ensembles are
effectively coupled via a flux qubit. We show that the coupling strengths
between flux qubits and spin ensembles can reach the strong and even
ultrastrong coupling regimes by either engineering the hybrid structure in
advance or tuning the excitation frequencies of spin ensembles via external
magnetic fields. When extending the hybrid structure to an array with equal
coupling strengths, we find that in the strong-coupling regime, the hybrid
array is reduced to a tight-binding model of a one-dimensional bosonic lattice. In the
ultrastrong-coupling regime, it exhibits quasi-particle excitations
separated from the ground state by an energy gap. Moreover, these
quasi-particle excitations and the ground state are stable under a certain
condition that is tunable via the external magnetic field. This may provide
an experimentally accessible method to probe the instability of the system.
\end{abstract}

\pacs{ 03.67.Ac, 42.50.Dv, 85.25.Cp, 76.30.Mi}
\maketitle


\section{\label{sec:level1}INTRODUCTION}

As an important subfield in quantum information, quantum simulation~\cite{1}
has attracted increasing interest, and considerable advancements were achieved
both theoretically~\cite{2,3,4,5,6,7} and experimentally~\cite{8,9,10,11,12}%
. Recently, hybrid quantum systems~\cite%
{13,14,15,16,17,18,19,20,21,22,23,24n,25n} have been strongly recommended
for quantum simulation because they can combine two distinct advantages of
the subsystems: the tunability of artificial atoms such as quantum circuits,
and the long coherence times of atoms or spins. Also, strong and tunable
coupling between two subsystems can be realized via either direct~\cite%
{19,24,25,26} or indirect~\cite{28,29,30} coupling schemes.

Among various hybrid quantum systems, the one consisting of superconducting
qubits and nitrogen-vacancy (NV) centers in a diamond~crystal \cite{28,29}
has become a topic of current interest. This hybrid quantum system has the
merits of high tunability and scalability in superconducting qubits, and long
coherence times and stable energy levels in NV centers. In addition, the
magnetic coupling between superconducting qubits and NV centers can be
stronger than that between NV centers and a transmission line resonator \cite%
{28} by three orders of magnitude. These distinct advantages make this
hybrid system a good candidate for simulating the abundant features of
many-body systems. For instance, a hybrid quantum architecture composed of
inductively coupled flux qubits, where a NV-center ensemble is placed on top
of each qubit loop, was proposed \cite{31} to simulate a Jaynes-Cummings
(JC) lattice. Using this hybrid quantum system, it is possible to
investigate the quantum phase transition between the localized and
delocalized phases in a Bose-Hubbard-like model. Because the flux qubits and
the NV centers can be tuned by external magnetic fields, the JC lattice
simulated using them is simpler and more tunable than those realized using
coupled cavities~\cite{32,33,34,35,36}.

In this paper, we study hybrid quantum systems consisting of spin ensembles
and superconducting flux qubits, where each spin ensemble is also realized
using the NV centers in a diamond crystal. Different from the hybrid quantum
system in Ref.~\cite{31}, a flux qubit is placed in between two NV-center
ensembles, and every two nearest-neighbor spin ensembles are effectively
coupled via this flux qubit. The coupling strength between flux qubits and
spin ensembles can be tuned to reach the strong- and even ultrastrong-coupling regimes by either engineering the hybrid structure in advance or
tuning the excitation frequencies of spin ensembles via the external
magnetic fields. These hybrid quantum systems can be used to simulate the
systems of coupled bosons, including the demonstration of bilinear coupling
among the bosons. Moreover, these hybrid systems have the advantage of
scalability, so they can be extended to construct hybrid arrays to simulate
one-dimensional~(1D) bosonic lattices with tunable coupling strengths. In the strong-coupling
regime, the hybrid array is simply reduced to the tight-binding model of a
1D bosonic lattice. In the ultrastrong-coupling regime, the hybrid array
exhibits more interesting properties. For instance, because of the bilinear
coupling in this regime, it can exhibit quasi-particle excitations that have
an energy gap from the ground state. Moreover, it is found that these
quasi-particle excitations and the ground state are stable under a certain
condition that is tunable via the external magnetic field. This may provide
an experimentally accessible method to probe the instability of the system.

The paper is organized as follows. In Sec.~II, we describe the proposed
hybrid quantum system consisting of two flux qubits and three adjoining spin
ensembles. The effective Hamiltonian for the spin ensembles is derived in
Sec.~III by adiabatically eliminating the degrees of freedom of the flux
qubits in the dispersive regime. In Sec.~IV, the proposed hybrid quantum
system is extended to a hybrid array, so as to simulate 1D bosonic lattices
with tunable coupling strengths. The behavior of this system in both the
strong- and the ultrastrong- coupling regimes is studied. Finally, discussions
and conclusions are given in Sec.~V.

\section{\protect\bigskip THE PROPOSED HYBRID QUANTUM SYSTEM}

We start from the simple structure consisting of three spin ensembles, with
every two nearest-neighbor ensembles coupled by a flux qubit [see Fig.~\ref%
{fig1}(a)]. We use NV centers in a diamond crystal as a spin ensemble \cite%
{22}. Each flux qubit is realized by a superconducting loop interrupted with
three Josephson junctions and biased by a static external magnetic field
perpendicular to the loop~\cite{37}. This superconducting qubit can have a
superposition state of clockwise and counterclockwise persistent currents in
the loop. In contrast to the previous proposals~\cite{28,29}, where a
NV-center ensemble is placed inside the loop of a flux qubit, we place a
flux qubit in between two NV-center ensembles in order to achieve effective
coupling of the NV-center ensembles and scalable hybrid structures.

\begin{figure}[tbp]
\includegraphics[width=8.5cm]{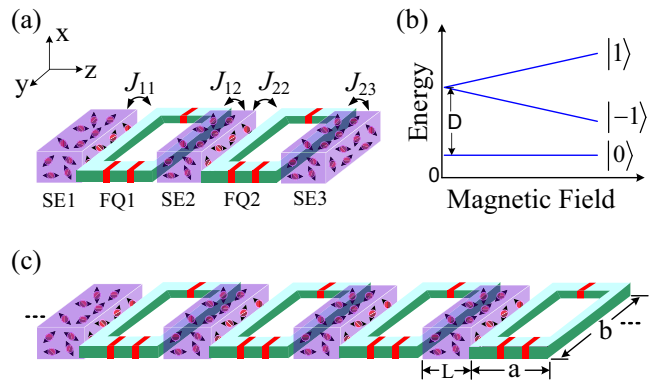}
\caption{(Color online) (a) Schematic diagram of the proposed hybrid quantum
system. Among the three spin ensembles, every two nearest-neighbor spin
ensembles are coupled by a superconducting flux qubit. The qubit loop is on
the y-z plane and perpendicular to the x direction. (b) The Zeeman splitting
for the spin-1 sublevels $m_{\mathrm{s}}=0,\pm 1$ of the electronic ground
state in a NV center. (c) Hybrid array consisting of spin ensembles and flux
qubits, which is a periodic extension of the hybrid quantum system in (a).}
\label{fig1}
\end{figure}

For a NV center, the spin-1 triplet sublevels of its electronic ground state
with $m_{\mathrm{s}}=0$ and $m_{\mathrm{s}}=\pm 1$ have a zero-field
splitting [see Fig.~\ref{fig1}(b)]. Because the strain-induced splitting is
negligible compared to the Zeeman splitting, the electronic ground state of
a single NV center can be described by~\cite{38}
\begin{equation}
H_{\mathrm{NV}}=DS_{\mathit{z}}^{2}+g_{e}\mu _{B}\mathbf{B}\cdot \mathbf{S},
\end{equation}%
where $D$ is the zero-field splitting of the electronic ground state, $%
\mathbf{S}$ is the usual spin-1 operator, $g_{e}$ is the g-factor, and $\mu
_{B}$ is the Bohr magneton. For convenience, we set the crystalline axis of
the NV center as the $z$ direction. As shown in Fig.~\ref{fig1}(b), by
introducing a weak external magnetic field $B_{\mathit{z}}^{\mathrm{ext}}$
along the $z$ direction, an additional Zeeman splitting between $m_{s}=\pm 1$
sublevels occurs. Thus, a two-level quantum system with sublevels $m_{s}=0$
and $-1$ can be separated from the other levels. In the subspace of this
two-level system, the Hamiltonian of a single NV center can be reduced to
(we set $h=1$)
\begin{equation}
H_{\mathrm{NV}}=\frac{1}{2}\nu _{s}\tau _{z},
\end{equation}%
where $\nu _{s}=D-g_{e}\mu _{B}B_{\mathit{z}}^{\mathrm{ext}}$ is the energy
difference between the lowest two sublevels with $m_{s}=0$ and $-1$, and $%
\mathbf{\tau }\equiv (\tau _{x},\tau _{y},\tau _{z})$ denote the Pauli
operators with the two basis states corresponding to the lowest two
sublevels.

The subsystem of two flux qubits has a Hamiltonian as follows~\cite{39}
\begin{equation}
H_{\mathrm{FQ}}=\overset{2}{\underset{i=1}{\sum }}\frac{1}{2}(\varepsilon
_{i}\sigma _{z}^{(i)}+\lambda _{i}\sigma _{x}^{(i)})+M_{12}\sigma
_{z}^{(1)}\sigma _{z}^{(2)},
\end{equation}%
where the first two terms involve two isolated flux qubits and the third one
is the interaction Hamiltonian of the two flux qubits coupled via a mutual
inductance $m_{12}$ defined by $M_{12}=m_{12}I_{p}^{(1)}I_{p}^{(2)}$, with $%
I_{p}^{(i)}$ being the persistent current of the $i$th flux qubit; $%
\varepsilon _{i}=2I_{p}^{(i)}(\Phi ^{(i)}-\Phi _{0}/2)$ is the energy bias
of the $i$th flux qubit (where $\Phi ^{(i)}$ is the applied static magnetic
flux, and $\Phi _{0}$ the flux quantum), $\lambda _{i}$ is the tunneling
energy, and $\boldsymbol{\sigma }\equiv (\sigma _{x}^{(i)},\sigma
_{y}^{(i)},\sigma _{z}^{(i)})$ denote the Pauli operators of the $i$th flux
qubit. To reduce the effect of the flux noise, the external static
magnetic field of each flux qubit is biased at the degeneracy point with $%
\varepsilon _{i}=0$, so that the transition frequency is $\nu _{qi}=\lambda
_{i}.$

Note that the two persistent-current states of each flux qubit can produce
an additional static magnetic field. This magnetic field leads to couplings
between the flux qubit and its two neighboring NV centers. In the following
discussions, we use the eigenstates of $\sigma _{x}^{(i)}$ as the new basis
states with $\sigma _{x}^{(i)}\rightarrow \sigma _{z}^{(i)}$, and $\sigma
_{z}^{(i)}\rightarrow \sigma _{x}^{(i)}$. Then, the Hamiltonian of the
hybrid quantum system in Fig.~\ref{fig1}(a) can be written as~\cite{21,31}
\begin{eqnarray}
H\! &\!=\!&\!\overset{2}{\underset{i=1}{\sum }}\frac{1}{2}\nu _{qi}\sigma
_{z}^{(i)}+M_{12}\sigma _{x}^{(1)}\sigma _{x}^{(2)}+\underset{j=1,2,3}{\sum }%
\overset{n}{\underset{m=1}{\sum }}\frac{1}{2}\nu _{s}^{(m)}\tau _{z,j}^{(m)}
\notag \\
&&\!+\overset{n}{\underset{m=1}{\sum }}[J_{11}^{(m)}(\tau _{1,+}^{(m)}+\tau
_{1,-}^{(m)})\sigma _{x}^{(1)}+J_{12}^{(m)}(\tau _{2,+}^{(m)}+\tau
_{2,-}^{(m)})\sigma _{x}^{(1)}  \notag \\
&&\!+J_{22}^{(m)}(\tau _{2,+}^{(m)}+\tau _{2,-}^{(m)})\sigma
_{x}^{(2)}+J_{23}^{(m)}(\tau _{3,+}^{(m)}+\tau _{3,-}^{(m)})\sigma
_{x}^{(2)}],~~~~  \label{4}
\end{eqnarray}%
where $J_{ij}^{(m)}\equiv \frac{1}{\sqrt{2}}g_{e}\mu _{B}B_{ij}^{(m)}$ is
the coupling strength between the $i$th flux qubit and the $m$th NV center
in the $j$th spin ensemble, with $B_{ij}^{(m)}$ $($ $i=1,2;$ $j=1,2,3$ $)$
being the corresponding magnetic field in the $x$ direction generated by the
$i$th flux qubit acting on the $m$th NV center in the $j$th spin ensemble.
The central spin ensemble (i.e., the second one) experiences the magnetic
fields generated by both the left and right neighboring flux qubits.

To describe the collective excitation of each spin ensemble, one can define~
\begin{equation}
s_{j}^{\dagger }=\frac{1}{J_{ij}}\underset{m=1}{\overset{n}{\sum }}%
J_{ij}^{(m)}\tau _{j,+}^{(m)},
\end{equation}%
where
\begin{equation}
J_{ij}=\sqrt{\sum_{m}|J_{ij}^{(m)}|^{2}}.
\end{equation}%
In the condition of large $n$ and low excitations for each spin ensemble, $%
s_{j}^{\dagger }$ and $s_{j}$ satisfies the bosonic commutation relation $%
[s_{j},s_{j}^{\dagger }]\approx 1$ (see, e.g., \cite{31,40,41}). Using these
collective operators, the total Hamiltonian in Eq.~(\ref{4}), can be
rewritten as
\begin{equation}
H=H_{\mathrm{FQ}}+H_{\mathrm{SE}}+H_{\mathrm{SE-FQ}}
\end{equation}%
with%
\begin{eqnarray}
H_{\mathrm{FQ}}\! &\!=\!&\!\frac{1}{2}\nu _{q1}\sigma _{z}^{(1)}+\frac{1}{2}%
\nu _{q2}\sigma _{z}^{(2)}+M_{12}\sigma _{x}^{(1)}\sigma _{x}^{(2)},  \notag
\\
H_{\mathrm{SE}}\! &\!=\!&\!\nu _{s1}s_{1}^{\dagger }s_{1}+\!\!\nu
_{s2}s_{2}^{\dagger }s_{2}+\nu _{s3}s_{3}^{\dagger }s_{3},  \label{8} \\
H_{\mathrm{SE-FQ}}\! &\!=\!&\!J_{11}(s_{1}^{\dagger }+s_{1})\sigma
_{x}^{(1)}+J_{12}(s_{2}^{\dagger }+s_{2})\sigma _{x}^{(1)}  \notag \\
&&+J_{22}(s_{2}^{\dagger }+s_{2})\sigma _{x}^{(2)}+J_{23}(s_{3}^{\dagger
}+s_{3})\sigma _{x}^{(2)}.  \notag
\end{eqnarray}

\begin{figure}[tbp]
\includegraphics[width=9cm]{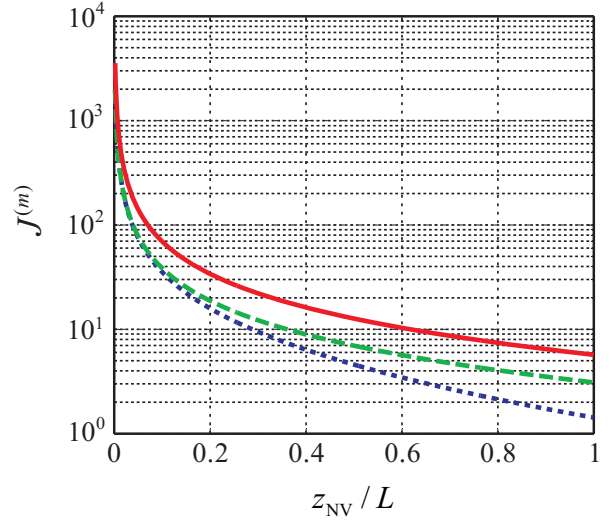}
\caption{(Color online) The separation ($z_{\mathrm{NV}}$) dependence of the
coupling strength $J^{(m)}$ between the flux qubit and the $m$th spin in the
NV-center ensemble, which is located on the symmetric line of the rectangular
loop in the $z$ direction. Here $I_{p}^{(i)}=0.5~\protect\mu\mathrm{A}$, and
$a=b=1~\protect\mu\mathrm{m}$ for the (blue) dotted curve; $I_{p}^{(i)}=0.5~%
\protect\mu\mathrm{A}, a=2~\protect\mu\mathrm{m}$, and $b=10~\protect\mu%
\mathrm{m}$ for the (green) dashed curve; and $I_{p}^{(i)}=0.9~\protect\mu%
\mathrm{A}, a=2~\protect\mu\mathrm{m}$, and $b=50~\protect\mu\mathrm{m}$ for
the (red) solid curve.}
\label{fig2}
\end{figure}

Below we estimate the coupling strength $J_{ij}$. The superconducting loop
of each flux qubit can be designed as, e.g., a rectangular loop with length $%
b$ and width $a$, where the narrow side with $a$ is along the $z$ direction
and the nearest-neighbor flux qubits are separated by $L$, i.e., the width
of the diamond crystal placed in between two adjoining flux qubits [see Fig.~%
\ref{fig1}(c)]. Therefore, we can denote the position of a NV center located
on the symmetric line along the $z$ direction of the nearest-neighbor
rectangular loops as $z_{\mathrm{NV}}$, where $0\leq z_{\mathrm{NV}}<L$.
According to the Biot-Savart law, the magnetic field generated by the left
flux qubit acting on this NV center can be written as
\begin{eqnarray}
B(z_{\mathrm{NV}})\! &\!=\!&\!\frac{\mu _{0}I_{p}}{\pi }\Big\{\frac{1}{\sqrt{%
(b/2)^{2}+(a+z_{\mathrm{NV}})^{2}}}  \notag \\
&&\!\times \Big[\frac{z_{\mathrm{NV}}+a}{b}+\frac{b}{4(z_{\mathrm{NV}}+a)}%
\Big]  \notag \\
&&\!-\frac{1}{\sqrt{(b/2)^{2}+z_{\mathrm{NV}}^{2}}}\Big(\frac{b}{4z_{\mathrm{%
NV}}}+\frac{z_{\mathrm{NV}}}{b}\Big)\Big\},~~~~
\end{eqnarray}%
where $\mu _{0}$ is the magnetic permeability. The corresponding coupling $%
J^{(m)}(z_{\mathrm{NV}})=\frac{1}{\sqrt{2}}g_{e}\mu B^{(m)}(z_{\mathrm{NV}})$%
, produced by the left flux qubit on this NV center, is shown in Fig.~2.
Here we choose the magnetic field $\bar{B}$ at\ $z_{\mathrm{NV}}/L=0.5$ as
the average magnetic field~\cite{21,29} to estimate the coupling produced by
the left qubit on its adjoining NV-center ensemble, i.e., $J=\sqrt{\frac{n}{2%
}}g_{e}\mu \bar{B}.$ From Fig.~\ref{fig2}, it can be seen that the coupling
for a single NV center decreases slightly when $z_{\mathrm{NV}}$ shifts from
$z_{\mathrm{NV}}/L=0.5$ to approaching $1$ (i.e., close to the right flux
qubit), but increases drastically when $z_{\mathrm{NV}}$ changes from $z_{%
\mathrm{NV}}/L=0.5$ to approaching $0$ (i.e., close to the left flux qubit).
Thus, it is expected that the actual coupling produced by the flux qubit on
its adjoining NV-center ensemble should be larger than $J$.

To estimate the value of the coupling $J$, we choose the experimentally
accessible density of NV centers as $3\times 10^{6}~\mu \mathrm{m}^{-3}$~%
\cite{42}, the height of the diamond crystal as $5~\mu \mathrm{m}$, and the
thickness of the superconducting loop as $60~\mathrm{nm}$~\cite{28}. For
persistent current $I_{p}^{(i)}=$ $0.5~\mu \mathrm{A}$, $a=b=1~\mu \mathrm{m}
$, and {$L=0.5~\mu \mathrm{m}$}, we have $J\approx \sqrt{n}J^{(m)}\sim 13~%
\mathrm{MHz}$. When $a$ is increased to $2~\mu \mathrm{m}$ and $b$ to $%
10~\mu \mathrm{m}$, $J\sim 60~\mathrm{MHz}$. This reaches the strong
coupling regime since the coupling strength produced by the flux qubit on
its adjoining NV-center ensemble is larger than the decoherence rates of
both the flux qubit ($\gamma _{\mathrm{FQ}}\sim 1~\mathrm{MHz})$ and the NV
center ensemble ($\gamma _{\mathrm{SE}}\sim 10~\mathrm{MHz}$~\cite{29}).
When $I_{p}^{(i)}$ is increased to $I_{p}^{(i)}=$ $0.9~\mu \mathrm{A}$~\cite%
{43}, and $b$ to $50~\mu \mathrm{m}$, the coupling $J$ is strengthened to $%
J\sim 250~\mathrm{MHz}$, which can be comparable to the excitation frequency
$\nu _{s}$ of the NV-center ensemble because $\nu _{s}\equiv D-g_{e}\mu
_{B}B_{\mathit{z}}^{\mathrm{ext}}$ can be tuned small via the external
magnetic field. This corresponds to the ultrastrong coupling regime. As for
the inductive coupling between the nearest-neighbor flux qubits, because
they are separated by $\sim 1~\mu \mathrm{m},$ it is as weak as $M_{12}\sim
1-10~\mathrm{MHz}$~\cite{44}. Therefore, compared to the coupling between
the flux qubit and its adjoining NV-center ensemble, the inductive coupling
can be neglected in the Hamiltonian (\ref{8}).

\section{\protect\bigskip EFFECTIVE HAMILTONIAN OF THE SPIN ENSEMBLES}

We rewrite the Hamiltonian $\mathrm{(7)}$ as
\begin{equation}
H=H_{\mathrm{0}}+H_{\mathrm{I}},
\end{equation}%
where the free part $H_{\mathrm{0}}$ is
\begin{equation}
H_{\mathrm{0}}=\frac{1}{2}\nu _{q1}\sigma _{z}^{(1)}+\frac{1}{2}\nu
_{q2}\sigma _{z}^{(2)}+\nu _{s1}s_{1}^{\dagger }s_{1}+\nu
_{s2}s_{2}^{\dagger }s_{2}+\nu _{s3}s_{3}^{\dagger }s_{3},
\end{equation}%
and the interaction part $H_{\mathrm{I}}$ is
\begin{eqnarray}
H_{I}\! &\!=\!&\!J_{11}(\sigma _{-}^{(1)}s_{1}^{\dagger }+\sigma
_{-}^{(1)}s_{1}+\mathrm{H.c.})  \notag \\
&&+J_{12}(\sigma _{-}^{(1)}s_{2}^{\dagger }+\sigma _{-}^{(1)}s_{2}+\mathrm{%
H.c.})  \notag \\
&&\!+J_{22}(\sigma _{-}^{(2)}s_{2}^{\dagger }+\sigma _{-}^{(2)}s_{2}+\mathrm{%
H.c.})  \notag \\
&&\!+J_{23}(\sigma _{-}^{(2)}s_{3}^{\dagger }+\sigma _{-}^{(2)}s_{3}+\mathrm{%
H.c.}).~~~~
\end{eqnarray}%
We consider the large-detuning case with $\Delta _{ij}\gg J_{ij}$ , where $%
\Delta _{ij}\equiv \nu _{qi}-\nu _{sj}>0$ ($i=1,2;$ $j=1,2,3$). In such a
case, the flux qubits can be regarded as being kept in their ground states
and only virtual excitations can occur. Therefore, we can use the Fr\"{o}%
hlich-Nakajima transformation~\cite{45,46,47} to adiabatically eliminate the
degrees of freedom of flux qubits to obtain an effective coupling between
the nearest-neighbor spin ensembles. This unitary transformation $U=\exp
(-V) $ requires the anti-Hermitian operator $V=-V^{\dagger }$ to satisfy
\begin{equation}
H_{\mathrm{I}}+[H_{\mathrm{0}},V]=0.
\end{equation}%
This gives rise to an effective Hamiltonian, up to second order, as
\begin{equation}
H_{\mathrm{eff}}=UHU^{\dagger }=H_{\mathrm{0}}+\frac{1}{2}[H_{\mathrm{I}%
},V]+O(J^{3}),
\end{equation}%
where the anti-Hermitian operator $V$ has the following form:
\begin{eqnarray}
V\! &\!=\!&\!A_{1}(\sigma _{-}^{(1)}s_{1}^{\dag }-\sigma
_{+}^{(1)}s_{1})+A_{5}(\sigma _{-}^{(1)}s_{1}-\sigma
_{+}^{(1)}s_{1}^{\dagger })  \notag \\
&&\!+A_{2}(\sigma _{-}^{(1)}s_{2}^{\dag }-\sigma
_{+}^{(1)}s_{2})+A_{6}(\sigma _{-}^{(1)}s_{2}-\sigma
_{+}^{(1)}s_{2}^{\dagger })  \notag \\
&&\!+A_{3}(\sigma _{-}^{(2)}s_{2}^{\dag }-\sigma
_{+}^{(2)}s_{2})+A_{7}(\sigma _{-}^{(2)}s_{2}-\sigma
_{+}^{(2)}s_{2}^{\dagger })  \notag \\
&&\!+A_{4}(\sigma _{-}^{(2)}s_{3}^{\dag }-\sigma
_{+}^{(2)}s_{3})+A_{8}(\sigma _{-}^{(2)}s_{3}-\sigma
_{+}^{(2)}s_{3}^{\dagger }),~~~~
\end{eqnarray}%
with the coefficients given by%
\begin{eqnarray}
A_{1}\! &\!=\!&\!\frac{J_{11}}{\Delta _{11}},~~A_{2}=\frac{J_{12}}{\Delta
_{12}},~A_{3}=\frac{J_{22}}{\Delta _{22}},~~A_{4}=\frac{J_{23}}{\Delta _{23}}%
,  \notag \\
A_{5} &=&\!\frac{J_{11}}{\Lambda _{11}},~~A_{6}=\frac{J_{12}}{\Lambda _{12}}%
,~~A_{7}=\frac{J_{22}}{\Lambda _{22}},~~A_{8}=\frac{J_{23}}{\Lambda _{23}}%
.\!~~~~~~~
\end{eqnarray}%
Here $\Lambda _{ij}\equiv \nu _{qi}+\nu _{sj}$ ($i=1,2;$ $j=1,2,3$)$.$

Because the coefficients $A_{l}$ ($l=1$ to $8$) are small in the
large-detuning case, the higher-order terms of the Fr\"{o}hlich-Nakajima
transformation can be dropped out and only the second-order term $[H_{%
\mathrm{I}},V]$ needs to be kept. Moreover, owing to the separate design of
the nearest-neighbor flux qubits in our approach, the diagonal term of each
flux qubit remains respectively at its original expectation value in the
adiabatic approximation, i.e., $\sigma _{z}^{(i)}\longrightarrow
\left\langle \sigma _{z}^{(i)}\right\rangle =-1,$ with $i=1,2.$ After
eliminating the degrees of freedom of each flux qubit, the effective
Hamiltonian can be obtained as
\begin{eqnarray}
H_{\mathrm{eff}}\! &&\!=\nu _{1}^{\prime }s_{1}^{\dag }s_{1}+\nu
_{3}^{\prime }s_{3}^{\dag }s_{3}+\nu _{2}^{\prime }s_{2}^{\dag
}s_{2}-g_{11}(s_{1}^{\dagger }s_{1}^{\dagger }+s_{1}s_{1})  \notag \\
&&\!-g_{22}(s_{2}^{\dagger }s_{2}^{\dagger
}+s_{2}s_{2})-g_{33}(s_{3}^{\dagger }s_{3}^{\dagger }+s_{3}s_{3})  \notag \\
&&\!-g_{12}(s_{1}s_{2}^{\dagger }+s_{2}s_{1}^{\dagger }+s_{2}^{\dagger
}s_{1}^{\dagger }+s_{1}s_{2})  \notag \\
&&\!-g_{23}(s_{2}s_{3}^{\dagger }+s_{3}s_{2}^{\dagger }+s_{3}^{\dagger
}s_{2}^{\dagger }+s_{2}s_{3}),  \label{17}
\end{eqnarray}%
where the parameters are given by
\begin{eqnarray}
\nu _{1}^{\prime }\! &=&\nu _{s1}-2g_{11},~~\nu _{3}^{\prime }=\nu
_{s3}-2g_{33},  \notag \\
\nu _{2}^{\prime } &=&\nu _{s2}-J_{12}^{2}\Big(\frac{1}{\Delta _{12}}+\frac{1%
}{\Lambda _{12}}\Big)-J_{22}^{2}\Big(\frac{1}{\Delta _{22}}+\frac{1}{\Lambda
_{22}}\Big),  \notag \\
g_{11}\! &\!=\!&\!\!\frac{J_{11}^{2}}{2}\Big(\frac{1}{\Delta _{11}}+\frac{1}{%
\Lambda _{11}}\Big),~~g_{33}=\frac{J_{23}^{2}}{2}\Big(\frac{1}{\Delta _{23}}+%
\frac{1}{\Lambda _{23}}\Big),  \notag \\
g_{22}\! &\!=\!&\frac{J_{12}^{2}}{2}\Big(\frac{1}{\Delta _{12}}+\frac{1}{%
\Lambda _{12}}+\frac{1}{\Delta _{22}}+\frac{1}{\Lambda _{22}}\Big),
\label{18} \\
g_{12}\! &\!=\!&\!\frac{J_{11}J_{12}}{2}\Big(\frac{1}{\Delta _{12}}+\frac{1}{%
\Lambda _{12}}+\frac{1}{\Delta _{11}}+\frac{1}{\Lambda _{11}}\Big),  \notag
\\
g_{23}\! &\!=\!&\!\frac{J_{22}J_{23}}{2}\Big(\frac{1}{\Delta _{23}}+\frac{1}{%
\Lambda _{23}}+\frac{1}{\Delta _{22}}+\frac{1}{\Lambda _{22}}\Big).  \notag
\end{eqnarray}%
Here $\nu _{si}^{\prime }$ is the effective excitation frequency of the $i$%
th spin ensemble, and $g_{12}$ ($g_{23}$) is the effective coupling strength
between the first (second) and second (third) spin ensembles. Therefore, we
achieve an effective coupling between the nearest-neighboring spin
ensembles, where each spin ensemble behaves like a boson. In Eq.~(\ref{17}),
there are bilinear terms $s_{i}^{\dagger }s_{i}^{\dagger }+s_{i}s_{i}$ ($%
i=1,2,3$), which involve one-mode squeezing within the same bosons. Also,
there are bilinear terms $s_{2}^{\dagger }s_{1}^{\dagger }+s_{1}s_{2}$ and $%
s_{3}^{\dagger }s_{2}^{\dagger }+s_{2}s_{3}$, which involve two-mode
squeezing between the three different bosons. These terms become important
in the ultrastrong coupling regime with $g_{ij}\sim \nu _{j}^{\prime }$ ($%
i=1,2;j=1,2,3$).

As a special case, we further consider the strong coupling regime, i.e., $%
\gamma _{\mathrm{FQ}},\gamma _{\mathrm{SE}}\ll J_{ij}\ll \Delta _{ij}$ and $%
g_{ij}\ll \nu _{j}^{\prime }$. In this case, the rotating-wave approximation
can be applied, and the effective Hamiltonian is reduced to
\begin{eqnarray}
H_{\mathrm{eff}}\! &\!=\nu _{1}^{\prime }\!&\!s_{1}^{\dagger }s_{1}+\nu
_{2}^{\prime }s_{2}^{\dagger }s_{2}+\nu _{3}^{\prime }s_{3}^{\dagger }s_{3}
\notag \\
&&\!-g_{12}(s_{2}^{\dagger }s_{1}+s_{1}^{\dag }s_{2})-g_{23}(s_{3}^{\dagger
}s_{2}+s_{2}^{\dagger }s_{3}),
\end{eqnarray}%
where $\nu _{i}^{\prime }$, $g_{12}$, and $g_{23}$ are given in Eq.~(\ref{18}%
). As compared with Eq.~(\ref{17}), the effective Hamiltonian is now reduced
to a JC form.

\section{HYBRID ARRAY}

As an extension of the hybrid structure in Fig.~\ref{fig1}(a), we now
propose a hybrid array in Fig.~\ref{fig1}(c). Here we consider a periodic
system with $\nu _{sl}=\nu _{s}$, $g_{l,l+1}=g$, and the periodic boundary
condition. Similar to Eq.~(\ref{17}), the effective Hamiltonian of the
hybrid array can be written as
\begin{eqnarray}
H\! &\!=\!&\!(\nu _{s}-2g)\underset{j}{\overset{}{\sum }}s_{j}^{\dagger
}s_{j}-g\underset{j}{\overset{}{\sum }}(s_{j}^{\dagger }s_{j}^{\dagger
}+s_{j}s_{j}  \notag \\
&&+s_{j}^{\dagger }s_{j+1}+s_{j+1}^{\dagger }s_{j}+s_{j}^{\dagger
}s_{j+1}^{\dagger }+s_{j+1}s_{j}),  \label{20}
\end{eqnarray}%
where
\begin{equation}
g=J^{2}\Big(\frac{1}{\Delta }+\frac{1}{\Lambda }\Big),
\end{equation}%
with $\Delta =\nu _{q}-\nu _{s},$ and $\Lambda =\nu _{q}+\nu _{s}.$

In order to diagonalize the Hamiltonian (\ref{20}), instead of using the
site operator $s_{j}^{\dagger }(s_{j})$ acting on the $j$th spin ensemble,
we employ the wave-vector operators $b_{k}^{\dagger }(b_{k})$ acting on all
ensembles\ in the array:

\begin{eqnarray}
s_{j}^{\dagger }\! &\!=\!&\!\frac{1}{\sqrt{N}}\underset{k}{\overset{}{\sum }}%
e^{-ik\cdot j}b_{k}^{\dagger },  \notag \\
s_{j}\! &\!=\!&\!\frac{1}{\sqrt{N}}\underset{k}{\overset{}{\sum }}e^{ik\cdot
j}b_{k}^{{}},
\end{eqnarray}%
where $N$ is the number of spin ensembles in the array. The new collective
operators present themselves like a wave excitation of bosons in the hybrid
array. We Fourier transform the bosonic operators to convert the Hamiltonian
to
\begin{align}\nonumber
H=\underset{k}{\sum }&\left\{\left[(\nu _{s}-2g(1+\cos k)\right]b_{k}^{\dagger}b_{k}\right.\\&\left.-g(1+e^{ik})
(b_{k}^{\dagger }b_{-k}^{\dagger }+b_{-k}b_{k})\right\},
\label{23}
\end{align}%
with $-\pi \leq k<\pi $. Because there are relations $b_{k}^{\dagger
}b_{-k}^{\dagger }=b_{-k}^{\dagger }b_{k}^{\dagger }$ and $%
b_{-k}b_{k}=b_{k}b_{-k}$ for bosons$,$ the Hamiltonian~(\ref{23}) can be
rewritten as
\begin{align}\nonumber
H=\underset{k}{\overset{}{\sum }}&\left\{\left[\nu _{s}-2g(1+\cos
k)\right](b_{k}^{\dagger }b_{k}+b_{-k}^{\dagger }b_{-k})  \notag \right.\\&\left.-2g(1+\cos k)(b_{k}^{\dagger }b_{-k}^{\dagger }+b_{-k}b_{k})\right\},
\end{align}%
where $0\leq k<\pi ,$ i.e., the wave vector $k$ is now confined in half of
the first Brillouin zone.

\begin{figure}[tbp]
\includegraphics[width=7.5cm]{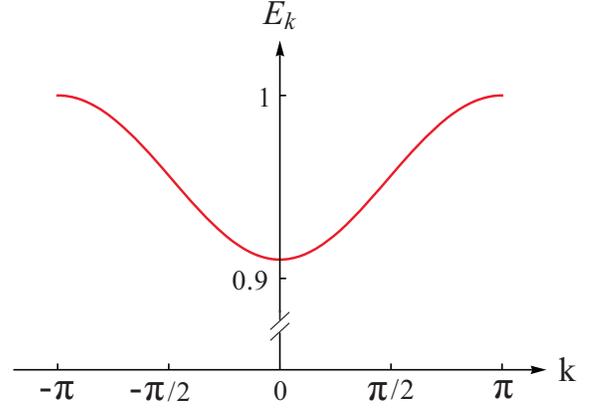}
\caption{(Color online) Dispersion relation for the quasi-particle energy $%
E_{k}$ in $\mathrm{Eq.~(27)}$, where $J=0.25~\mathrm{GHz}$, $\protect\omega %
_{q}=6~\mathrm{GHz}$ and $\protect\omega _{s}=1~\mathrm{GHz}$. Note that an
energy gap occurs, which separates the quasi-particle excitations from the
ground state. }
\label{fig3}
\end{figure}

Furthermore, we apply the Bogoliubov transformation:
\begin{eqnarray}
b_{k}\! &\!=\!&\!\mu _{k}\alpha _{k}-\nu _{k}\alpha _{-k}^{\dagger },  \notag
\\
b_{k}^{\dagger }\! &\!=\!&\!\mu _{k}^{\ast }\alpha _{k}^{\dagger }-\nu
_{k}^{\ast }\alpha _{-k},  \notag \\
b_{-k}\! &\!=\!&\!\mu _{k}\alpha _{-k}-\nu _{k}\alpha _{k}^{\dagger },
\notag \\
b_{-k}^{\dagger }\! &\!=\!&\!\mu _{k}^{\ast }\alpha _{-k}^{\dagger }-\nu
_{k}^{\ast }\alpha _{k},
\end{eqnarray}%
where $|\mu _{k}|^{2}-|\nu _{k}|^{2}=1.$ It is clear that $\alpha _{k}$ and $%
\alpha _{k}^{\dagger }$ also satisfy the bosonic commutation relations: $%
[\alpha _{k},\alpha _{k^{\prime }}^{\dagger }]=\delta _{kk^{\prime }},$ and $%
[\alpha _{k},\alpha _{k^{\prime }}]=[\alpha _{k}^{\dagger },\alpha
_{k^{\prime }}^{\dagger }]=0.$ Then, the Hamiltonian is diagonalized to%
\begin{equation}
H=\underset{k}{\overset{}{\sum }}E_{k}(\alpha _{k}^{\dagger }\alpha
_{k}+\alpha _{-k}^{\dagger }\alpha _{-k})+E_{g},
\end{equation}%
where the quasi-particle energy is
\begin{equation}
E_{k}=\sqrt{\nu _{s}^{2}-4\nu _{s}g(1+\cos k)},  \label{27}
\end{equation}%
and the ground-state energy is
\begin{equation}
E_{g}=\underset{k}{\overset{}{\sum }}\Big\{\sqrt{\nu _{s}^{2}-4\nu
_{s}g(1+\cos k)}-[\nu _{s}-2g(1+\cos k)]\Big\}.
\end{equation}%
Here the parameters are required to satisfy
\begin{equation}
\nu _{s}\geqslant 8g.  \label{29}
\end{equation}%
In the opposite limit with $\nu _{s}<8g,$ the eigenmodes contain imaginary
solutions, and the system is unstable. This indicates that when the
frequency detuning is fixed, the coupling strength $J$ between the flux
qubits and the spin ensembles should be bounded even in the ultrastrong
regime, so as to achieve a stable quantum system. The dispersion relation
for the quasi-particle energy $E_{k}$ with the given parameters is shown in Fig.~\ref{fig3}, where an energy gap separates the quasi-particle excitations from
the ground state. Experimentally, one can tune $\nu _{s}\equiv D-g_{e}\mu
_{B}B_{\mathit{z}}^{\mathrm{ext}}$ via the external magnetic field $B_{%
\mathit{z}}^{\mathrm{ext}}$ on the NV centers in order to satisfy the
condition in Eq.~(\ref{29}).

Meanwhile, in the limit of $\nu _{s}\gg g$, the counter rotating terms in $H$
can be neglected and the effective Hamiltonian of the hybrid array is
reduced to
\begin{equation}
H=(\nu _{s}-2g)\underset{j}{\overset{}{\sum }}s_{j}^{\dagger }s_{j}-g%
\underset{j}{\sum }(s_{j}^{\dagger }s_{j+1}+s_{j+1}^{\dagger }s_{j}).
\end{equation}%
This Hamiltonian corresponds to the tight-binding model of bosons on a 1D
lattice. We can diagonalize the Hamiltonian as
\begin{equation}
H=\underset{k}{\sum }E_{k}b_{k}^{\dagger }b_{k},
\end{equation}%
where
\begin{equation}
E_{k}=\nu _{s}-2g(1+\cos k).  \label{32}
\end{equation}%
This dispersion relation agrees with Eq.~(\ref{27}), in the limit of $\nu
_{s}\gg g$:
\begin{eqnarray}
E_{k} &=&\nu _{s}\Big[1-\frac{4g}{\nu _{s}}(1+\cos k)\Big]^{1/2}  \notag \\
&\approx &\nu _{s}\Big[1-\frac{2g}{\nu _{s}}(1+\cos k)\Big]  \notag \\
&=&\nu _{s}-2g(1+\cos k).
\end{eqnarray}%
Similar to the energy bands of non-interacting electrons in a crystal, the
dispersion relation in Eq.~(\ref{32}), determines an energy band of the 1D
bosonic crystal.

\section{DISCUSSION AND CONCLUSION}

\bigskip Compared with the scheme composed solely of superconducting qubits,
our proposed hybrid system can be more easily tuned to the ultrastrong-coupling regime. Experimentally, this can be conveniently achieved by tuning
the applied magnetic field to increase the ratio of the coupling strength
between the flux qubit and the NV-center ensemble to the excitation
frequency of the spin ensemble. As given in Sec.~II, for the experimentally
accessible parameters $I_{p}=$ $0.9~\mu \mathrm{A}, a=2~\mu \mathrm{m}$, and $%
b=50~\mu \mathrm{m}$, the coupling strength $J$ between the flux qubit and
the NV-center ensemble can be as strong as $J\sim 250~\mathrm{MHz}$. If the
frequency of the NV-center ensemble is tuned, for example, to $\nu _{s}\sim
1~\mathrm{GHz}$ via the external magnetic field, one can reach an
ultrastrong-coupling regime with $J/\nu _{s}\sim 25\%$. The corresponding
ratio of the effective coupling strength $g$ between bosons to the
excitation frequency of the spin ensemble can reach $g/\nu _{s}\sim 12\%$.
However, in the well-established scheme composed solely of superconducting
qubits~\cite{44}, where two flux qubits are coupled via an additional
large-detuned flux qubit, the ratio of the effective coupling strength
between the two flux qubits to the single-qubit frequency is about $1.5\%$.
Obviously, it is much smaller than the achievable ratio $g/\nu _{s}\sim 12\%$
in our hybrid system.

Coherence times in superconducting qubits have been steadily increasing over
the past decade, with the coherence times in excess of $100$ $\mu \mathrm{s}$
for the transmon qubit in a 3D cavity {\cite{48n,49n}}. Due to the large
capacitance shunted to the qubit, the effect of the charge noise on the
qubit is greatly suppressed. Also, it is this large shunt capacitance that
yields a strong coupling between the qubit and the cavity. However, it is
hard to use only one qubit to both simultaneously and strongly couple two
cavities, due to the difficulty in circuits design. Also, it could be a
similar case for the proposed 1D array of resonators coupled by either
superconducting-ring couplers or dc-SQUIDs~\cite{50n}, because it is usually not easy to achieve ultrastrong couplings for a superconducting-ring coupler (dc-SQUID) simultaneously coupled to two resonators. In contrast, as discussed above, our
proposed hybrid system can be more conveniently tuned to the ultrastrong-coupling regime by just tuning the applied magnetic field to increase the
ratio of the qubit-spin coupling to the excitation frequency of the spin
ensemble.

In a coupled system, it is difficult to directly calculate the relaxation or
decoherence times of the subsystems because they depend on the nature of the
environment coupled to each subsystem. However, they can be estimated from
some experimental observations. For instance, in a recent experiment on the
hybrid system consisting of a flux qubit and a NV-center ensemble, it was
estimated from the observed decay time of the quantum oscillations that,
when the flux qubit was in resonance with the NV centers, the relaxation
times were $T_{\mathrm{1,qb}}\sim 150$ $\mathrm{ns}$ for the flux qubit and $%
T_{\mathrm{1,NV}}\gg 10$ $\mu \mathrm{s}$ for the NV centers \cite{29}. In
our proposed hybrid system, the flux qubit is kept in the ground state, so
the coherence times of the NV centers should be even longer since the
designed off-resonance of the flux qubit to the NV centers will suppress the
decoherence on the NV centers induced by the relaxation of the flux qubit.
In fact, in our proposed hybrid system, the flux qubit is largely detuned
from the NV centers. Typically, the frequency $\nu _{q}$ of the flux qubit
can be designed as $5$-$10$ $\mathrm{GHz}$. In the experiments of
superconducting qubits, the temperature can be as low as $T\sim 10$ $\mathrm{%
mK}$, which corresponds to $k_{B}T\sim 0.1$ \textrm{GHz }$\ll \nu _{q}$.
Thus, even a lossy flux qubit can indeed be kept in the ground state at such
a low temperature, and its relaxation-induced decoherence on the NV centers
is then suppressed.

{In conclusion}, we have studied a hybrid quantum system consisting of spin
ensembles and superconducting flux qubits. Each spin ensemble is realized
using the NV centers in a diamond crystal, and every two nearest-neighbor
spin ensembles are effectively coupled by adiabatically eliminating the
degrees of freedom of the flux qubit placed in between these two spin
ensembles. The coupling strengths between flux qubits and spin ensembles can
be tuned into strong- and even ultrastrong- coupling regimes by either
engineering the hybrid structure in advance or tuning the excitation
frequencies of spin ensembles via the external magnetic fields. As an
elementary structure, three effectively coupled spin ensembles can be used
to simulate the system of coupled bosons, especially for demonstrating
bilinear coupling between bosons in the ultrastrong-coupling regime.
Moreover, due to the advantage of scalability, this structure can be
extended to construct hybrid arrays to simulate 1D bosonic lattices with
tunable coupling strengths. In the strong-coupling regime, the hybrid array
can be simply reduced to the tight-binding model on a 1D bosonic lattice.
However, in the ultrastrong-coupling regime, the hybrid array exhibits more
interesting properties. Because of the bilinear coupling in this regime, it
can exhibit quasi-particle excitations that have an energy gap from the
ground state. Moreover, these quasi-particle excitations and the ground
state are stable under a certain condition that is tunable via the external
magnetic field.

\section*{ACKNOWLEDGMENTS}

This work is supported by the National Natural Science Foundation of China
Grant No.~91121015, the National Basic Research Program of China Grant
No.~2014CB921401, and the NSAF Grant No.~U1330201. L.T. is supported by the
National Science Foundation under Awards No. 0956064 and No. 0916303.

\end{document}